# Machine learning methods for nanolaser characterization

Darko Zibar, Molly Piels, Ole Winther, Jesper Mørk and Christian Schaeffer

*Abstract:* **Nanocavity lasers, which are an integral part of an on-chip integrated photonic network, are setting stringent requirements on the sensitivity of the techniques used to characterize the laser performance. Current characterization tools cannot provide detailed knowledge about nanolaser noise and dynamics. In this progress article, we will present tools and concepts from the Bayesian machine learning and digital coherent detection that offer novel approaches for highly-sensitive laser noise characterization and inference of laser dynamics. The goal of the paper is to trigger new research directions that combine the fields of machine learning and nanophotonics for characterizing nanolasers and eventually integrated photonic networks**.

INTRODUCTION

Optical technologies have been dominating the long distance information superhighways of the internet [1-3]. However, power efficient light based communication has also started migrating to ever-shorter distances and will eventually move onto silicon chips in the near future. The next step in the evolution of the optical technologies is towards realization of a photonic network on a silicon CMOS chip. The envisioned applications of on-chip networks are in the fields of data communication and optical sensing. For a photonic network on a chip, the density requirement, large modulation bandwidth and ultra-low power consumption are critical issues [4]. The reason is that several thousand lasers, photodetectors and switches must be integrated and modulated on a single chip. For the successful realization of the on-chip photonic network, the noise properties of the integrated lasers are of critical importance. Generally speaking, as the on-chip components get smaller, the available output power becomes limited and the role of quantum noise becomes increasingly important. This translates into more stringent requirements on the measurement sensitivity.

Sources based on photonic crystal nanocavities are excellent candidates for on-chip light sources as they offer low-power operation and high modulation speeds [5-7]. Recent theoretical results even suggest that nanolasers may achieve modulation bandwidths of the order of several hundred GHz, not limited by relaxation oscillations, and acceptable noise performance [7-8].

However, for building on-chip photonics networks based on nano cavity laser sources, there are still many challenges that need to be addressed. Firstly, accurate noise characterization of the nano cavity sources is crucial for any of the applications [4-9]. A significant challenge in the characterization of nanolasers is that the output power is in the micro and nano Watt regime [4-9]. The low output power limits the available signal-to-noise ratio and imposes stringent requirements on the sensitivity of the measurement techniques employed.

Secondly, the estimation and characterization of the parameters that govern the dynamics of the nanocavities is crucial in order to establish accurate simulation models for system performance evaluations and for optimizing the laser and chip design [9-11]. While, for conventional semiconductor lasers, the rate of spontaneous emission into the laser mode, the so-called spontaneous emission beta-factor, is of the order of $10^{-5}$, beta approaches 1 for nanolasers [11] yielding a very smooth threshold curve and in general complicating the extraction of laser parameters and the separation of signal and noise contributions. The majority of current methods for estimating laser parameters rely on small signal analysis. The small signal analysis, although extremely useful for qualitative understanding of the operation of nano cavity lasers, is not generally applicable when the laser is operating close to threshold or large-signal modulated above the threshold. Moreover, current state-of-the-art laser characterization methods only provide a point estimate, i.e. single value used as a best guess of an unknown parameter. Ideally, from the laser designer point of view, it would be useful to quantify the confidence in the point estimates intervals and thereby provide confidence intervals of the estimated parameters.

Finally, due to a complex interdependency between the signal and noise within the nano cavity, a mathematical model that describes the dynamics of the nanolaser or photonic integrated circuit may be very difficult to obtain [9-10]. Instead, it is therefore desirable to build a framework that learns the model of the nanolaser from the actual data. For instance, it has recently been demonstrated that minute fabrication-disorder has a decisive influence on the laser threshold, an effect that comes about due to the strong dispersion of photonic crystals which enhances light-matter interaction via slow-light effects [12]. Performing model inference of the nano lasers dynamics from the measurement data, it is possible to get a better representation of the actual operation of the nanolaser and capture effects not described by the analytical models.

In general, performing noise characterization, parameter estimation and model inference is also challenging for other nanocavity systems where the power is extremely low, making it very difficult to assess the system dynamics, such as cavity opto-mechanical systems [13-15].

The field of Bayesian machine learning combined with digital coherent detection, which has recently had a large impact in the field of optical communication, offers an attractive approach to address the above mentioned challenges [16-17]. The introduction of the digital coherent detection has allowed for the reconstruction of the optical field in the discrete time (digital) domain, opening opportunities for novel characterization methods. State-of-the-art digital coherent detectors have been demonstrated to operate up to 100 GHz of analog bandwidth and sampling rates in the range of 160 Gs/s, opening opportunities for new approaches to perform ultra-fast characterizations [22-24].

The field of Bayesian machine learning offers powerful tools to perform highly sensitive measurements, infer parameters from observable data, which may be sparse and noisy; infer plausible models to describe the observed data; and use the inferred models to make predictions based on the past input data [18-21]. However, the field of Bayesian machine learning spans over a large number of topics and it is therefore essential to select the appropriate methods that are beneficial for nanolaser characterization.

By employing digital coherent detection, the observable data represents a sequential time series of the noisy laser output signal. The observable data will contain latent or hidden variables which are the dynamic and static parameters of the nanolaser. We therefore need to choose Bayesian machine learning methods that are capable of inferring plausible models of the latent variables (dynamic and static laser parameters) from the observable noisy data. To address that task, we need to employ machine learning methods that rely on Bayesian inference in state-space models, i.e. Bayesian filtering. Bayesian filtering employing state-space models also allows for the inclusion of the underlying physical principles of the nanolasers which is very useful from the nanolaser operational point of view.

In conclusion, the aim of this paper is to address machine learning methods, based on the nonlinear state-space Bayesian filtering, and link them to the current problems for nanolaser characterization. Finally, it will be briefly reviewed how the proposed methods can also be applied for characterization of cavity opto-mechanical systems.

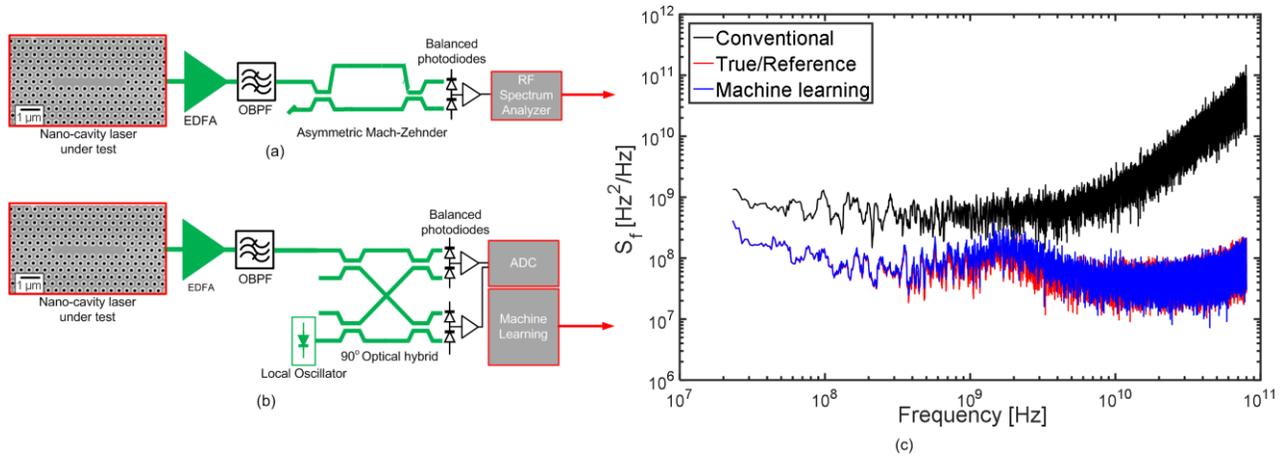

**Figure 1:** (a) Conventional approach for laser noise characterization. (b) Laser noise characterization employing digital coherent detection and machine learning for offline processing. (c) *Simulations:* Frequency noise spectra of the nanolaser using the conventional and machine learning, in combination with digital coherent detection, based approach. EDFA: erbium doped fibre amplifier, OBPF: Optical bandpass filter, ADC: analogue-to-digital-converter.

## Main principles

The conventional and dominant approach for laser noise characterization is to measure the noise spectra using a pair of balanced photodiodes and electrical spectrum analyzer as illustrated in Fig. 1(a). For frequency noise spectrum measurements, the phase fluctuations must first be converted to intensity fluctuations by a frequency discriminator, such as an asymmetric mach-zehnder interferometer. This approach is severely limited at low output powers by noise from the photodiodes and the amplifiers that are typically required and thereby does not provide the required sensitivity for the nanolaser characterization. This is illustrated in Fig. 1(c), where the frequency noise spectrum obtained using the conventional measurement technique significantly differs from the true/reference frequency noise spectrum. The frequency noise spectrum in Fig. 1(c) is dominated by noise and it is not possible to resolve the shape of the laser frequency noise spectrum and make conclusions about the laser dynamics. Finally, the conventional approach does not allow for the reconstruction of the detected optical field in the digital domain, which is necessary for model inference as explained in the next section.

Fig. 1(b) illustrates nanolaser characterization employing digital coherent detection in combination with machine learning. The main principle behind the digital coherent detection is to heterodyne the optical signal from the nanolaser, under the test, with the local oscillator (LO) laser and detect the in-phase and quadrature components of the beat signal by employing a 90 degree optical hybrid, a pair of balanced photodiodes and an analogue-to-digital-converter. The samples can then be stored and processed offline. The employed LO laser typically has very low intensity and phase noise. Current state-of-the-art lasers have linewidths below 100 Hz and relative intensity noise (RIN) less than 140 dBc/Hz [23]. The effectiveness of the proposed approach, combination of the digital coherent detection and machine learning methods employing Bayesian filtering, is illustrated in Fig. 1(c). It is observed that the obtained frequency noise spectrum closely resembles the true/reference frequency noise spectrum.

State-space models for nanolaser dynamics

Next, we will go into detail on how to apply Bayesian inference methods for laser noise and parameter characterization. For that purpose, we refer to Fig. 2(a). We will introduce the following coloring scheme. Blue is used for the measurable quantities and red for the unobservable quantities, see also Fig 2. To begin with, it is assumed that the dynamics of the nanolaser can be described by standard rate equations. The rate equations are a set of time-domain stochastic differential equations that describe the evolution of dynamic parameters such as the carrier density, $N(t)$, photon density, $S(t)$, and optical phase $\phi(t)$, and their relationship [25]. The rate equations are parametrized by static parameters that describe the laser dynamical behavior and are directly related to the laser's physical design. The static nano cavity laser parameters are: the mode confinement factor, $\Gamma$, the carrier density at the transparency, $N_t$, the photon lifetime, $\tau_p$, the fraction of the spontaneous emission coupled into the lasing mode, $\beta$, the active layer volume, $V$, the gain slope, $g_0$, the linewidth enhancement factor, $\alpha$, the gain compression factor, $\varepsilon$, and the injection efficiency $\eta_i$. Finally, the Langevin noise sources $\Gamma_N(t)$, $\Gamma_S(t)$ and $\Gamma_\phi(t)$ are Gaussian random processes with zero mean and correlation functions that are delta-correlated (white noise) [34].

The strengths of the Langevin noise sources are specified by the variances, $\sigma^2_N$, $\sigma^2_S$ and $\sigma^2_\phi$, respectively. For ease of notation, we will group all the unknown parameters, including the noise variances, into a vector of unknown static parameters $\theta$. The dynamic and the static nanolaser parameters are referred to as hidden or non-observable variables as they cannot be directly measured. Digital coherent detection allows for the reconstruction of the amplitude and phase of the optical field, $y(t)$, and is related to the dynamic and static parameters of the laser such as the photon density, $S(t)$, and optical phase, $\phi(t)$. This is illustrated in Fig 2(b). The quantity $y(t)$ represents the available measurements (dataset) and will contain static and dynamic laser parameters.

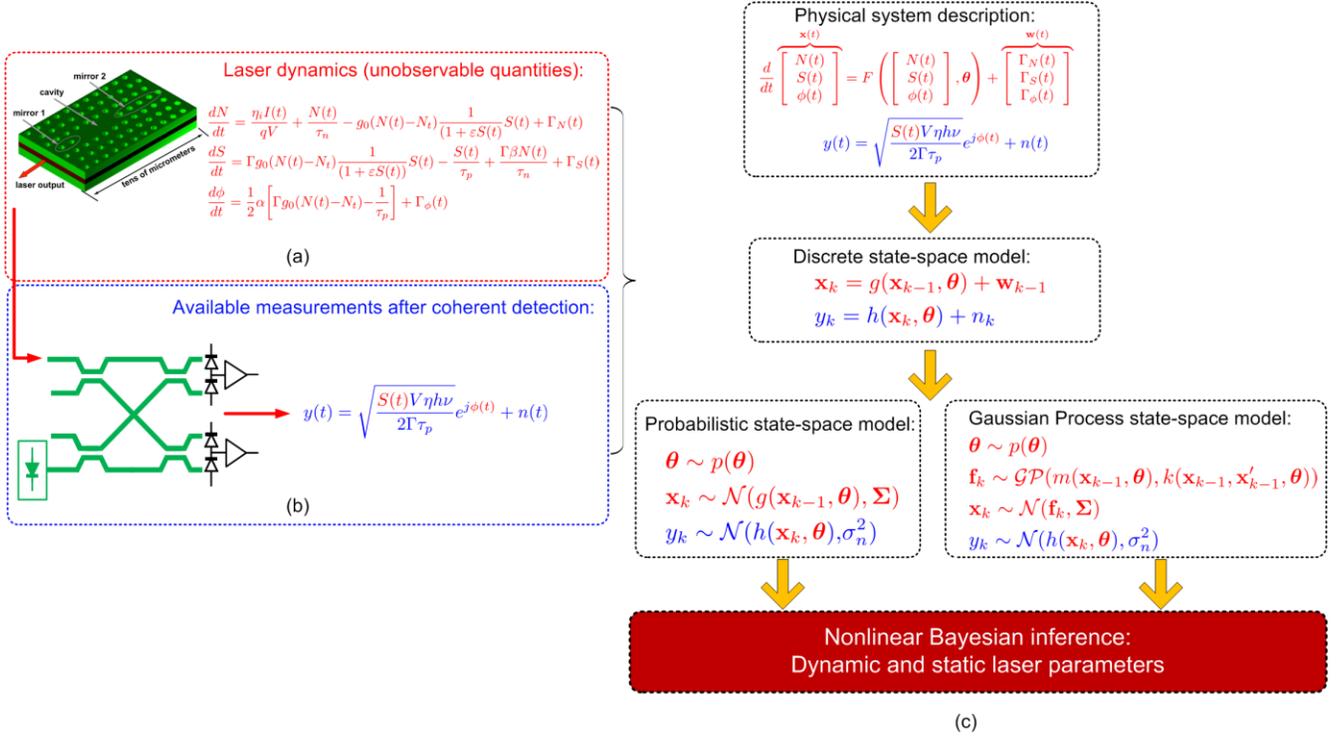

**Figure 2:** (a) Nano cavity laser and the corresponding rate equations. (b) Reconstructed optical field after digital coherent detection. $\eta_0$ is the differential quantum efficiency, $\nu$ is the unmodulated optical frequency and $h$ is Planck's constant. (c) State-space model combining rate equations and measurements for the static and dynamic parameter inference.

The measurements will be affected by the noise, *n(t),* which originates from various optical and electrical noise sources, so called measurement noise. The available measurements *y(t)* together with the stochastic differential rate equations form a nonlinear dynamical system and represent the available physical model of the entire system under consideration, see Fig. 2(c). The main idea is to use the available measurements *y(t)* and the known underlying structure of the rate equations to infer the dynamic and static parameters of the nano cavity laser. To perform dynamic and static parameter inference, it is convenient to express the measurements and rate equations in state-space form. The state-space model (SSM) offers a general and very powerful tool to learn, model, and analyze nonlinear dynamical systems. Since the measurement data will be available in sampled form after the analogue-to-digital-converter, only discrete state-space models are considered in this paper. There are several approaches to convert a time-continuous nonlinear dynamical system into a discrete time state-space model. The most common approach is to use a so-called Euler-Maruyama scheme [26].

Referring to Fig. 2(c), the discrete time state-space model consists of two equations: 1) the state equation which describes the evolution of state variables, *$x_k$*; and 2) the measurement equation is specified by observable variables $y_k$. The state variables $x_k$ are the time-varying parts of the system; for the considered case the state variables are the carrier density, photon density and the optical phase. The state equation consists of a deterministic part specified by the function $g()$ which encodes the dynamics of the rate equations and is parameterized by static parameter vector $\theta$, and the stochastic part specified by the Langevin noise sources. Using the dynamical system terminology, the stochastic part is referred to as process noise $w_k$. It should be noted that the measurement equation, $y_k$, and the rate-equations describing the state evolution are nonlinear, which makes both parameter inference and state tracking challenging.

From the Bayesian inference point of view, it is more convenient to express the deterministic state-space model in a probabilistic state-space model as illustrated in Fig. 2(c). The probabilistic state-space model will allow us to explore the knowledge, we already have, about the static cavity parameters as well as to quantify uncertainty in final parameter estimates. For instance, we may know the approximate range of values of the static parameters. This range and the corresponding mean can then be used to parameterize the prior probability density function of the static parameters $p(\theta)$. The first equation in the probabilistic state-space model describes that the static parameters are drawn from the prior probability density function, i.e. $\theta \sim p(\theta)$. Moreover, since the measurement noise is Gaussian, it means that the measurements, $y_k$, can be represented as samples drawn from a Gaussian distribution with mean $f(x)$ and variance, $\sigma^2_n$, specified by the measurement noise, i.e. $y_k \sim N(y_k|f(x_k),\sigma^2_n)$. Similarly, since the Langevin noise sources are Gaussian, the states $x_k$ can be represented as drawn from Gaussian distribution with mean $g(x_{k-1})$ and variance specified by the strength of the Langevin noise terms, i.e. $x_k \sim N(x_k|g(x_{k-1}),\sigma^2_v)$.

Until now, we have assumed that the underlying physical model of the laser has been known. In the case of the nanolaser, it was assumed that the laser dynamics were fully described by the rate equations specified in Fig. 2(a). However, in many cases due to the complexity of the interactions within the cavity and effects of structural disorder, the true system dynamics may not be known or may be only partially understood [7]. In that case, the Bayesian inference approach allows us to use all prior knowledge available to infer the uncaptured laser dynamics from the measurement data and thereby improve the measurement accuracy. We refer to this problem as model inference because we would like to infer the dynamical model of the laser from the measured data. In short, we would like to infer the mapping function *g()*.

In general, for the laser model inference is similar to the one for static parameter inference as we would also explore some prior knowledge about the laser cavity dynamics. However, in this case, in addition to inferring the dynamic and static parameters, we would also infer the dynamical model of the nano-cavity laser. For that particular case, the Gaussian Process state-space model needs to be employed [27-29], see Fig. 2(c).

In general, the Gaussian process is an effective technique to learn the complex mapping function. In brief, this means that given the input and output of the system denoted by: $y_k=f(x_k)+n_k$, the output values $y_k$ can be considered as drawn from a normal distribution where the mean, $m(x_k)$, and covariance, $k(x_k,x_{k'})$ are dependent on the input $x_k$ and $x_{k'}$. The mean and covariance functions $m(x_k)$ and $k(x_k,x'_k)$ are usually assumed known up to some parameters that determines the smoothness of the function $f(x_k)$. Typically, $m(x_k)$ will be based on the known part of the rate equations, derived from the small signal analysis, that describes the underlying system. However, since those equation may not completely describe the system, the uncaptured dynamics will be included though the covariance or kernel function $k(x_k,x'_k)$.

The Kernel function can take different functional forms such as: squared exponential, linear and Matern [25]. A model selection is needed to determine which kernel is the most appropriate one for a given dataset. The Kernels will typically be specified by parameters, so called hyper parameters, which then need to be inferred. In general, all the hyper-parameters and the static unknown parameters can be grouped in the vector theta $\theta$. Model selection can be performed either by investigating what choice of kernel function that has the largest marginal likelihood (which requires marginalization over $\theta$) or by computing the likelihood on a held-out test dataset.

## State-space models for laser dynamics

Once the state-space model has been formulated, we can proceed with Bayesian inference which includes learning of the static and dynamic laser parameter and the underlying model. The generic framework of the Bayesian inference is illustrated in Fig. 3. For simplicity, many details are omitted as the main purpose is to give an overview of the iterative scheme. For more detailed explanations the interested reader is referred to references [20-21, 30].

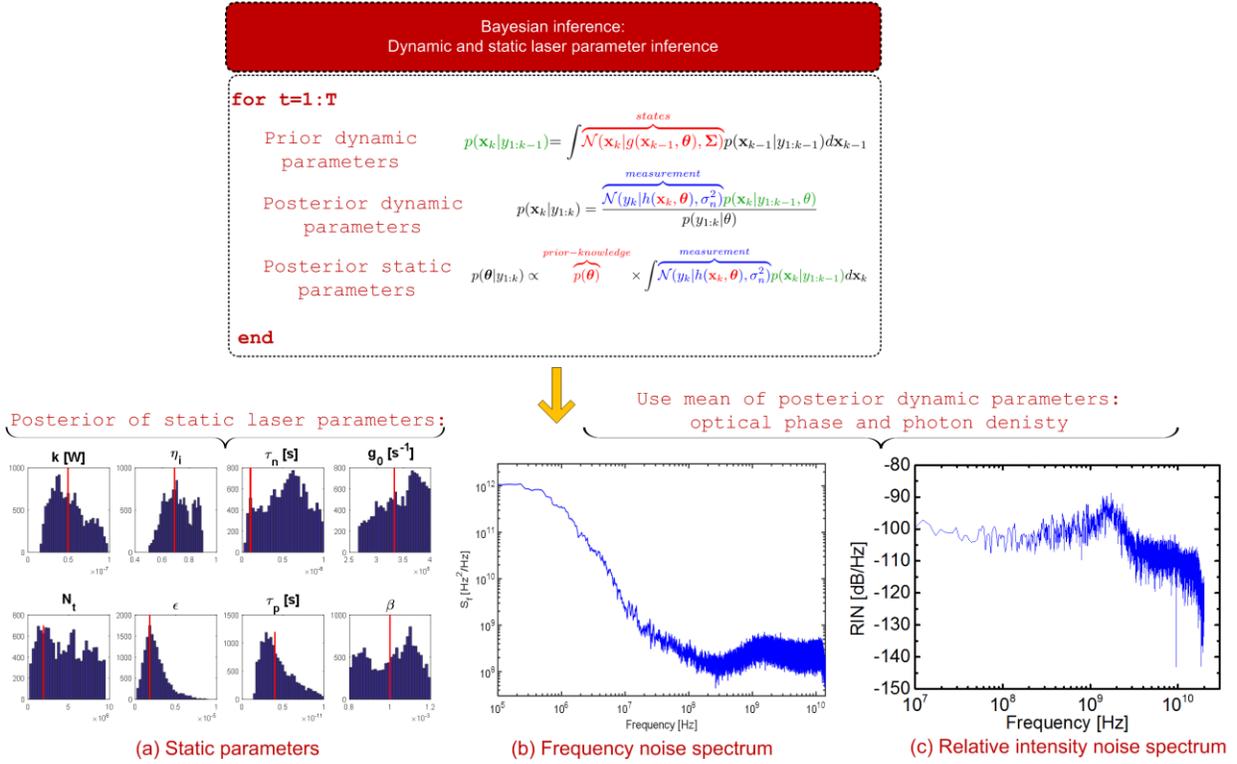

**Figure 3:** Application of Bayesian filtering framework for dynamic and static parameter estimation. (a) Histogram of the estimated static parameters, $\theta$, based on the simulation data. We have defined $k = \frac{V\eta h\nu}{2\Gamma\tau_p}$. The red line denotes the true values. The estimated dynamic parameters are photon density and optical phase which are then used to compute frequency and relative intensity noise spectra (b)-(c). The data is taken from the measurements reported in [33].

The objective of the nonlinear Bayesian inference is to iteratively infer the dynamic and static parameters, $x_k$ and $\theta$, respectively, from the measurement data $y_k$. More precisely, we would like to determine the probability density function of $x_k$ and $\theta$, given the measurement data $y_k$ for the finite time interval $k=[1:T]$. The probability density functions for $x_k$ and $\theta$, are denoted by $p(x_k|y_{1:T})$ and $p(\theta|y_{1:T})$. The probability density functions $p(x_k|y_{1:T})$ and $p(\theta|y_{1:T})$ encompass all the uncertainties associated with the parameters. The point estimates $x_k^{est}$ and $\theta^{est}$ at time $k$ are then obtained by computing the mean of the probability distributions $p(x_k|y_{1:T})$ and $p(\theta|y_{1:T})$, respectively.

In more technical terms, the prior (predictive) probability density function, $p(x_k|y_{1:k-1})$, given the measurements up to discrete time $k-1$ and candidate static parameters, is first computed. Then, once we have an available measurement at discrete time $k$ expressed by $p(y_k|x_k)$, the prior distribution is updated and we compute the posterior distribution $p(x_k|y_{1:k})$ and also we obtain a new estimate the static parameters. Moreover, the posterior probability distributions can be used to obtain a confidence interval of the extracted parameters by computing covariance of $x_k$ and $\theta$. In Fig. 3, the mean of the estimated optical phase and photon density from the measurements is used to obtain frequency and relative intensity noise spectra. The Bayesian inference problem, estimation of the mean of the optical phase and photon density is solved by employing particle and Kalman filtering. Moreover, we show the histogram of the estimated static parameters.

In general, solving the nonlinear Bayesian state inference problem of jointly estimating the static and dynamic parameters is a challenging task and there are no closed form analytical solutions. Current research efforts within the Bayesian machine leaning community are focusing on various iterative computational methods. One approach is to combine sequential Monte Carlo methods for state estimation with Monte Carlo Markov Chain methods for parameter estimation. Another approach is to consider the parameter estimation as an optimization problem and use approaches based on maximum likelihood approaches. It falls outside of the scope of this paper to address this methods and the interested reader is referred to the following references [31-32] and the references therein.

## Application to other systems

We have already proved the effectiveness of the nonlinear Bayesian inference methods and demonstrated, detailed frequency and relative intensity noise characterization of photonic crystal cavity nanolaser [17,33]. Furthermore, it has been shown that for the complex laser phase noise profiles similar to the ones that nano-cavity lasers may exhibit, successful traditional phase tracking algorithms fail, while successful phase tracking can be performed by employing nonlinear Bayesian inference methods that invokes the underlying laser physics [34-35]. This is a crucial point if nano-cavity lasers are going to employ phase modulation for information transport to enhance the overall information throughput. A related approach of using linear Bayesian inference methods has recently been applied to optimal state estimation of cavity opto-mechanical systems [15]. State estimation of a cavity opto- mechanical system is essential for optimal state control and verification. The proposed approach in reference [15] overcomes the limitation of conventional techniques and allows demonstration of real-time optimal state estimation for cavity opto-mechanical systems operating in arbitrary parameter regimes [15]. Similarly to the laser dynamics, the dynamics of the opto-mechanical architecture can also be described by time-domain Langevin equations. Moreover, the optical field from the opto-mechanical cavity can also be detected by employing coherent detection as shown in Fig. 4.

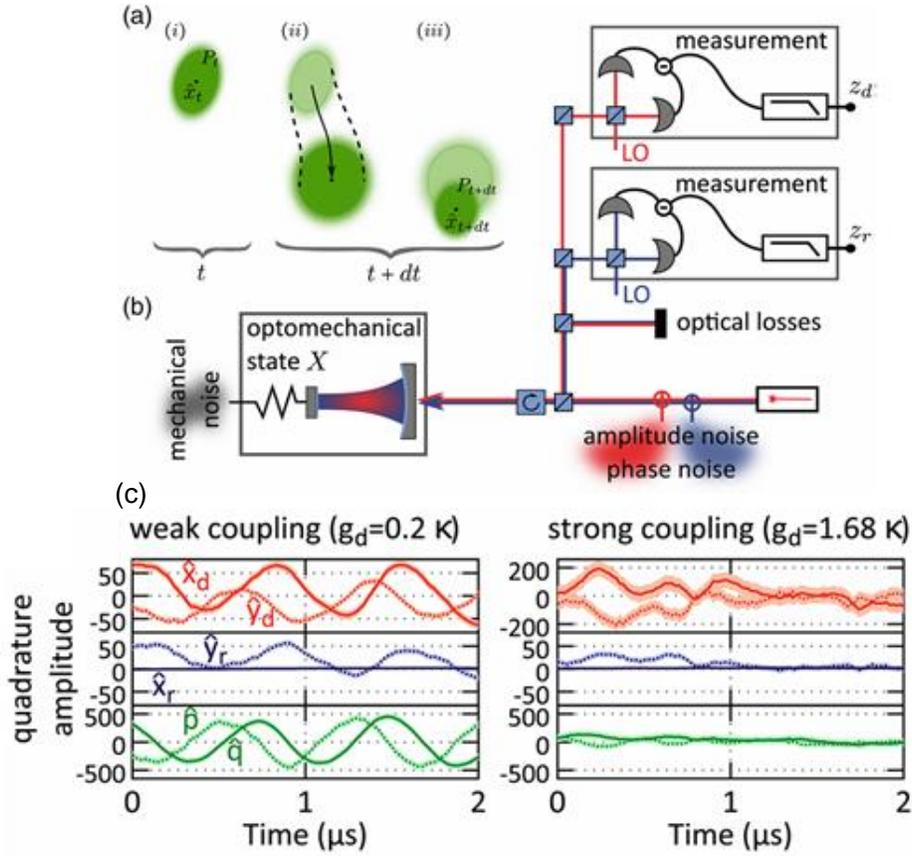

**Figure 4:** Figure taken from refernce [15]: Kalman filter for cavity optomechanical systems. (a) Working principle of the Kalman filter: (i) The conditional state is depicted by the green phase-space ellipse, which (ii) evolves in time according to the system dynamics. (iii) After a time dt, a Bayesian update is applied based on the measurement outcome to find the new conditional state. This procedure minimizes the mean-square estimation error, which makes the Kalman filter optimal for real-time state estimation. (b) Schematic of the experiment: The optomechanical cavity is driven by two laser beams each of which carry amplitude and phase noise. The mechanical motion is typically driven by Brownian noise. After their interaction with the cavity, the optical fields are detected by two independent homodyne measurements(signals zd and zr), which themselves are subject to optical losses and noise. Building an accurate Kalman filter requires appropriate modeling of all relevant noise sources. (c) Estimation of cavity states: optical amplitude, $y_i$, and phase, $x_i$, beam along with the mechanical position, q, and momentum quadrature, $p_i$.

The noisy measurements of the detected optical field are functions of the states of the opto-mechanical cavity and the previously described Bayesian inference techniques can then be applied. In reference [15], the authors use a linearized version of quantum Langevin equation, which can be related to the small signal analysis. Due to the linearized form of the quantum Langevin equations, linear Bayesian inference methods, which is equivalent to a Kalman filtering framework, is applied. As a final conclusion of the paper, the authors also state that Kalman filtering offers a significant performance advantage for classical and quantum control of cavity opto-mechanical systems. One of the extensions of the work presented in [15] would be to explore the nonlinear Bayesian inference methods which would allow avoidance of the potential limitations imposed by the linearization of the quantum Langevin equations and enhance static and dynamic parameter estimation.

CONCLUSION

Nano-cavity lasers are essential for the realization of an on-chip photonic network due to their low power consumption, large modulation bandwidth and small size. However, numerous challenges remain open in the detailed characterization and also modelling of these nano sources due to limited output power and complex cavity dynamics. To address these challenges, we need to look beyond current state-of-the-art characterization methods.

In this paper, it has been presented how techniques from Bayesian machine learning, more specifically nonlinear state-space Bayesian inference, can be used to perform highly sensitive noise characterization and model inference of the nano-cavity lasers. These techniques can be effectively used to obtain detailed noise characterization in terms of frequency and relative intensity noise, and also static nanolaser parameters. Even though we are still in the early stage of the development and there are still a lot more topics to be explored, these techniques present novel ways of not only studying nano-cavity lasers but also other nano-cavity systems such as cavity opto-mechanical systems.